\documentclass[12pt]{article}

\usepackage{amssymb}
\usepackage{amsmath}
\usepackage{latexsym}

\catcode`@=11
\renewcommand{\section}{\@startsection{section}{1}{0pt}{\medskipamount}
{\medskipamount}{\large\bf}}
\numberwithin{equation}{section}
\catcode`@=12

\def\beq{\begin{eqnarray}}    
\def\eeq{\end{eqnarray}}      

\def\pa{\partial}                       

\def\={\ =\ }

\begin{document}

\begin{center}

{\Large\bf Quantum  localization of Classical Mechanics}

\vspace{18mm}

{\large Igor A. Batalin$^{(a, b)}\footnote{E-mail:
batalin@lpi.ru}$\; and\;
Peter M. Lavrov$^{(b, c)}\footnote{E-mail:
lavrov@tspu.edu.ru}$
}

\vspace{8mm}

\noindent ${{}^{(a)}}$
{\em P.N. Lebedev Physical Institute,\\
Leninsky Prospect \ 53, 119 991 Moscow, Russia}

\noindent  ${{}^{(b)}}
${\em
Tomsk State Pedagogical University,\\
Kievskaya St.\ 60, 634061 Tomsk, Russia}

\noindent  ${{}^{(c)}}
${\em
National Research Tomsk State  University,\\
Lenin Av.\ 36, 634050 Tomsk, Russia}

\vspace{20mm}

\begin{abstract}
\noindent
Quantum localization of classical mechanics within the BRST-BFV and BV (or field-antifield)
quantization methods are studied. It is shown that a special choice of
gauge fixing functions (or BRST-BFV charge) together with the
unitary limit leads to Hamiltonian localization in the path integral
of the BRST-BFV formalism. In turn, we find that a special choice of gauge fixing
functions being proportional to extremals of an initial non-degenerate classical action
together with a very special solution of the classical master equation result in Lagrangian
localization in the partition function of the BV formalism.
\end{abstract}

\end{center}

\vfill


\\

\section{Introduction and Summary}

In Theoretical Physics, one considers  usually how to quantize complex
dynamical systems, possibly with nontrivial geometry of phase space or  constraints
of both classes, by observing precisely all the fundamental Physical Principles
\cite{DeWitt,Weinberg,HT,brs1,t}.
Although the general quantization problem  is of great importance, it is also very
interesting to realize what is the status of the classical dynamics from the point of
view of an equivalent special quantum theory. In other words, it is interesting to study, how to
formulate a version of a quantum theory capable, in a consistent way, to suppress
all quantum fluctuations  in a given dynamical system, so that the functional
path integral reduces to the delta - functional concentrated at the classical
trajectory, multiplied by the Jacobian of the argument of the delta - functional with
respect to the trajectory. That is exactly what we actually mean when saying about the
general quantum localization. There exist a rather numerous series of articles
where such a program has been realized and studied \cite{G,GRT1,GRT2,DG, CGN}.
One of the most known approaches is to introduce, in a Hamiltonian dynamics,
in addition to the usual Boson
time, the two Fermion time  components \cite{CGN}.
The linear coefficients of a series expansion
of the trajectory  in the new Fermion times become then a kind of
Hamiltonian ghosts, while the quadratic coefficients become Lagrange multipliers
yielding  just the delta - functional concentrated at the classical trajectory. In turn,
the Hamiltonian ghosts yield the Jacobian required  of the delta - functional.   Although
that approach by itself  looks very nice, it seems that the idea of two Fermion extra
times is rather artificial. On the other hand, in that scheme, the localization is
essentially discrete, in the sense that there is no "parameter"
interpolating continuously, as "switching on" quantum fluctuations.

In the present article, we suggest to use the
well-known BRST - BFV scheme invented originally for generalized canonical quantization
of relativistic dynamical systems with first - class constraints \cite{FVh1,BVh}.  We
suggest an unusual form for the BRST - BFV charge and gauge Fermion, and show that
there exists such a choice of the gauge - fixing functions that, in the unitary
limit, the path integral reproduces exactly the general Hamiltonian localization. In that
limit, the Lagrange multipliers and antighosts become dynamically passive, which leads
naturally  to the appearance of the set of the delta - functionals  necessary as to the
localization. In contrast to the genuine relativistic gauge theories, now the gauge -
fixing functions is intended to kill the dynamically active Lagrange multipliers to the gauge
fixing  itself, which looks a bit paradoxical, like an attempt of  taking one from a bog by
using the M\"{u}nchhausen trick. However, that is actually the case. In the approach
suggested, just the gauge - fixing function plays the role of an interpolation
"parameter" switching on quantum fluctuations.

Further, as we have known that the whole field - antifield formalism \cite{BV,BV1}
is derived from the BRST-BFV quantization scheme \cite{BF1,GGT1,GGT2,BT},
it appears quite natural
that there should be possible to describe the localization phenomenon just in the
language of the field - antifield formalism, as well. 
We
realize that program, by formulating a very specific  version of the classical master -
action. The action is a homogeneous linear function of the antifields, so that it
vanishes at zero values of the ones. There is no term of the zero order in the
antifields. The latter circumstance makes it possible for the unity matrix to stand for
the "gauge generator". Then, we also introduce a very specific version to the
gauge - fixing function, which is proportional to the extremal of the original
action. As a coefficient of that proportionality, we allow for an invertible
field-dependent metric. There also included the term proportional
to the Nakanishi - Lautrup
fields, with the coefficient being the delocalization parameter. As a result, at the
zero value of the delocalization parameter, we arrive at  the functional path
integral for the partition function, where the integration over the Nakanishi -
Lautrup - fields yield the delta - function of the extremal of the original action,
while the integrations over the ghosts and antighosts yield the corresponding Jacobian
correct. On the other hand, if the delocalization parameter is not zero,
there appears the Gaussian distribution of the classical extremal, instead of the
delta - function. In fact, as the metric and the delocalization parameter do enter
the functional path integral  via the gauge - fixing only, the whole  partition
function is independent of them, due to the standard general arguments \cite{BVh,BV,BV1,KT,BF}.

\section{Hamiltonian localization in the  BRST-BFV formalism}

Let $S$ be an original Hamiltonian action of some nonsingular dynamical
system,
\beq
\label{eq.1}
S = \int dt \left[ \frac{1}{2} Z^{A} \omega_{AB} \dot{Z}^{B} - H( Z ) \right],     
\eeq
where $Z^{A}$ are phase variables, $\varepsilon( Z^{A} ) = \varepsilon_{A}$,
$\omega_{AB}$ is a constant invertible symplectic metric, $H( Z )$
is a non-degenerate  original
Hamiltonian. The action (\ref{eq.1}) yields the standard equations of motion,
\beq
\label{eq.2}
\dot{Z}^{A} = \{ Z^{A} , H(Z) \}_{\omega} = \omega^{AB} \partial_{B} H( Z ).        
\eeq
Now, we would like to perform a localization to the equation of motion (\ref{eq.2})
via the following extended action $W$ ,
\beq
\label{eq.3}
W = \int dt \left[ P_{A} \dot{Z}^{A} + \Pi_{A} \dot{\Lambda}^{A} +
\bar{{\cal P}}_{A} \dot{C}^{A}
+ \bar{C}_{A} \dot{ \mathcal{P} }^{A} - \mathcal{H} \right],       
\eeq
in relativistic phase space of canonical  pairs with the values assigned of
the Grassmann  parities, $\varepsilon$, and ghost number, ${\rm gh}$,
original phase variables:
\beq
\label{eq.4}
( P_{A}, Z^{A} ), \quad \varepsilon( Z^{A} ) = \varepsilon( P_{A} ) = \varepsilon_{A},\quad
{\rm gh}( Z^{A} ) = - {\rm gh}( P_{A} ) = 0;                  
\eeq
dynamically active Lagrange multipliers:
\beq
\label{eq.5}
( \Pi_{A}, \Lambda^{A} ), \quad   \varepsilon( \Lambda^{A} ) = \varepsilon( \Pi_{A} )
= \varepsilon_{A},\quad
{\rm gh}( \Lambda^{A} ) = - {\rm gh}( \Pi_{A} )  = 0;      
\eeq
ghost phase variables:
\beq
\label{eq.6}
( \bar{ \mathcal{ P } }_{A}, C^{A} ),  \quad\varepsilon( C^{A} ) = \varepsilon(
\bar{ \mathcal{ P } }_{A} ) =
 \varepsilon_{A} + 1, \quad {\rm gh}( C^{A} ) = - {\rm gh}( \bar{ \mathcal{ P } }_{A} ) = 1;
\eeq
antighost phase variables:
\beq
\label{eq.7}
( \bar{C}_{A}, \mathcal{ P }^{A} ), \quad \varepsilon( \mathcal{ P }^{A} ) =
\varepsilon( \bar{C}_{A} ) =
\varepsilon_{A} + 1, \quad {\rm gh}( \mathcal{ P }^{A} ) = - {\rm gh}( \bar{C}_{A} ) = 1.
\eeq
The unitarizing extended Hamiltonian is given by
\beq
\label{eq.8}
\mathcal{ H } = \{ \Psi, \Omega \},      
\eeq
where gauge Fermion $\Psi$ and BRST charge $\Omega$ are
\beq
\label{eq.9}
\Psi = \bar{ \mathcal{ P } }_{A}  \Lambda^{A} + \bar{C}_{A} \chi^{A}, \quad
\varepsilon( \Psi ) = 1, \quad  {\rm gh}( \Psi ) = -1,      
\eeq
\beq
\label{eq.10}
\Omega = H \overleftarrow{ \partial_A} C^{A} + \Pi_{A} \mathcal{ P }^{A},\quad
\varepsilon(\Omega) = 1,\quad  {\rm gh}(\Omega) = 1,\quad  \{ \Omega, \Omega \} = 0,   
\eeq
where $\chi^{A}$ are just gauge - fixing functions, $\varepsilon( \chi^{A} ) =
\varepsilon_{A}$,  ${\rm gh}( \chi^{A} ) = 0$,
which may depend on all the phase variables, and
$\partial_{A}$ is everywhere a partial $Z^{A}$ - derivative.
The minimal form (\ref{eq.8}) of the unitarizing  Hamiltonian
$\mathcal{H}$ is directly related to the
possibility to use actually the so-called quantum/derived antibrackets
 \cite{K-S1,K-S2,BM1,BM2}  as to  describe dynamical systems with first-class constraints
(see \cite{BL} for details); see also \cite{M,M1} for the modified
description of the time evolution for general gauge systems.

By substituting (\ref{eq.9}), (\ref{eq.10}) into (\ref{eq.8}),
we get the following explicit form of the unitarizing extended Hamiltonian,
\beq
\label{eq.11}
\mathcal{ H }  = H \overleftarrow{\pa_A}\;\! \Lambda^{A} + \Pi_{A} \chi^{A}  +
\bar{C}_{A} \{ \chi^{A}, \Pi_{B} \mathcal{ P }^{B} \} +
\bar{C}_{A}\{ \chi^{A}, H \overleftarrow{ \pa_B} C^{B} \} +
\bar{ \mathcal{ P } }_{A} \mathcal{ P }^{A} .     
\eeq
All the Poisson brackets in (\ref{eq.8}) - (\ref{eq.11}) are
defined with respect to the canonical pairs in (\ref{eq.4}) -
(\ref{eq.7}). In terms of the extended action (\ref{eq.3}), the
partition function reads \beq \label{eq.12}
\mathcal{ Z } = \int D\Gamma  \exp\left\{\frac{i}{\hbar} W \right\},            
\eeq
where $D\Gamma$ is a trivial integration measure over all the phase variables
(\ref{eq.4}) - (\ref{eq.7}).

Due to the standard general arguments based on the Poisson - bracket
nilpotency of $\Omega$ in (\ref{eq.10}), the partition function (\ref{eq.12})
 is independent of the gauge
Fermion $\Psi$, and thereby of the gauge - fixing functions
$\chi^{A}$ \cite{FVh1}. On the other hand, for general $\chi^{A}$,
of course, the functional path integral (\ref{eq.12}) is not
localized to a solution to the classical equation of motion
(\ref{eq.2}), with the Jacobian of the corresponding delta
functional being compensated exactly.  We will show however that
there exists such a choice for $\chi^{A}$ that the required
localization is the case in the unitary limit of the path integral
in (\ref{eq.12}). So, let us define the unitary limit in the usual
way \cite{FVh1,FVh2,BVh,BV,BV1,BF1}.  As the partition function
(\ref{eq.12}) is independent of $\chi^{A}$, let us formally rescale
the latter \beq \label{eq.13}
\chi^{A}\quad  \rightarrow \quad \alpha^{ -1 } \chi^{A},            
\eeq
where $\alpha$ is a parameter. Then, let us make the following unimodular
change of integration variables in (\ref{eq.12}),
\beq
\label{eq.14}
\Pi_{A} \quad  \rightarrow \quad \alpha \Pi_{A},\quad
\bar{C}_{A} \quad  \rightarrow \quad \alpha \bar{C}_{A}.         
\eeq
Now, let us go to the limit $\alpha \rightarrow 0$. It is easy to see that
the structure of the Hamiltonian (\ref{eq.11}) is preserved under
the limit. The only result of the unitary limit in the extended
action $W$ is that the second and fourth kinetic terms in the
integrand in (\ref{eq.3}) become absent, so that the Lagrange multipliers
(\ref{eq.5}) and antighosts (\ref{eq.7}) become dynamically -
passive variables. It is now the case to choose a special  gauge -
fixing,
\beq
\label{eq.15}
\chi^{A} = \Lambda^{A} - P_{B}\; \omega^{BA}.               
\eeq
In the latter case, the $\Pi_{A}$ - integration yields the delta functional
$\delta[ \Lambda -P \omega]$, while the $\mathcal{P}^{A}$ - integration yields
the delta-functional $\delta[  \bar{
\mathcal{P} } +\bar{C} ]$.
By integrating out these delta - functionals, we have to
substitute
\beq
\label{eq.16}
\Lambda^{A} = P_{B} \;\omega^{BA},   \quad   \bar{\mathcal{ P } }_{A} = -
\bar{C}_{A},       
\eeq
so that the final action becomes
\beq
\label{eq.17}
W_{final} =  \int dt [ P_{A} ( \dot{ Z }^{A} - \omega^{AB} \pa_{B} H ) -
\bar{C}_{A} ( \dot{C}^{A}  -
\omega^{AC} \overrightarrow{\pa_C}\;\! H \;\!\overleftarrow{ \pa_B} C^{B} ) ].
\eeq
Here, the $P_{A}$ - integration yields the delta-functional
\beq
\label{eq.18}
\delta[ \dot{Z} - \{ Z, H \}_{\omega} ],         
\eeq
while the $\bar{C}_{A}$ and  $C^{B}$ - integrations yield exactly the Jacobian of
the argument in (\ref{eq.18}) with respect to $Z$. Thus, we have
reproduced exactly the result of
Hamiltonian localization.

Now, it is worth to compare the final action (\ref{eq.17}) to the known approach
based  on the use of the two extra time Fermionic
variables $\theta^{ a },  a = 1, 2$,  where  one  proceeds with the
two-parametric  superfield  action of the form \cite{DG,BB}
\beq
 \label{eq.19}
W'  =  \int dt  d^{ 2 }\theta \;[  \frac{ 1 }{ 2 }  Z^{A} \omega_{ AB } \dot{
Z }^{ B }  -  H( Z )  ],   \quad \omega_{ AB }  =  {\rm const}( Z ),      
\eeq
 where
\beq
 \label{eq.20}
Z^{A}( t, \theta^{ 1 }, \theta^{ 2 } )  =  Z^{A}_{0}( t )  +  \theta^{ a }
Z^{A}_{ a }( t )  +  \delta^{ 2 }( \theta )  Z^{ A }_{ 3 }( t ),   
\eeq
 is the component expansion of the two-parametric superfield,
\beq
 \label{eq.21}
&&d^{ 2 }\theta  =  d\theta^{ 1 } d\theta^{ 2 },   \quad  \delta^{ 2 }( \theta )  =
\frac{ 1 }{ 2 } \theta^{ a } \varepsilon_{ ab} \theta^{ b},   
\\
\label{eq.22} &&\int  d^{ 2 }\theta \;\delta^{ 2 }( \theta )  =  1,
\quad   \int d^{ 2 }\theta\;
\theta^{ a } \theta^{ b }  =  - \varepsilon^{ ab },   
\eeq
 while (\ref{eq.21}), (\ref{eq.22}) represent the two-parametric integration measure,
delta-function, and the normalization properties.
Due to (\ref{eq.20}) - (\ref{eq.22}), the action (\ref{eq.19})
rewrites in the component form
\beq
 \label{eq.23}
W' =  \int  dt \; [   Z^{ A }_{ 3 }  (  \omega_{AB}  \dot{ Z }^{ B }_{ 0 }  -
\partial_{ A } H( Z_{ 0 } )  )  +
\frac{ 1 }{ 2 } \varepsilon^{ ab }  (   Z^{A}_{ a } \omega_{ AB }
\dot{ Z }^{ B }_{ b }  -
 Z^{ A }_{ a } \overrightarrow{ \partial_A} \; \!H  \;\!\overleftarrow{ \partial_
B }  Z^{ B }_{ b }   )  ( -1 )^{ \varepsilon_{ B } }   ].   
\eeq
 It follows when comparing (\ref{eq.17}) to (\ref{eq.23})
 that the variables do identify as
\beq
 \label{eq.24}
Z^{ A }_{ 3 } \omega_{ AB }  =  P_{ B },  \quad   Z^{A}_{ 1} \omega_{ AB } =
\bar{ C }_{ B },  \quad  -  Z^{ B }_{ 2 } ( -1 )^{ \varepsilon_{ B } }  =  C^{
B}.     
\eeq

Notice that the explicitly $N=2$ supersymmetric form of the action
(\ref{eq.19}) reads \cite{Hull}
\beq
\label{eq.25}
W'  =  \int  dt  d^{2}\theta  [ -  \frac{ 1 }{ 4 }  D_{ a } Z^{ A }\;\! \omega_{
AB } \;\! g^{ ab }  D_{ b } Z^{ B }  (-1)^{ \varepsilon_{ B } }  -  H( Z ) ], 
\eeq
where covariant superderivatives are given by
\beq
\label{eq.26}
D_{ a }  =  \frac{ \partial }{ \partial \theta^{ a } }  +  g_{ ab }\;\! \theta^{
b } \frac{ \partial }{ \partial  t },    
\eeq
with
\beq
\label{eq.27}
g_{ ab }  =  g_{ ba }  =  {\rm const},  \quad    \varepsilon( g_{ ab } )  =  0, 
\eeq
being an invertible symmetric  constant metric, and $g^{ ab }$ being an
inverse to the latter.
Due to the supercommutation relations
\beq
\label{eq.28}
[  D_{ a },  D_{ b }  ]  =  2  g_{ ab }  \frac{ \partial }{ \partial  t }, 
\eeq
the explicitly $N = 2$  supersymmetric action (\ref{eq.25}) coincides exactly with
the one (\ref{eq.19}) after integration by part.

\section{Some possible generalizations}
We have demonstrated an interesting possibility of the Hamiltonian
generalized BRST-BFV formalism \cite{FVh1,BVh}, being a very
powerful quantization method of arbitrary dynamical systems with
constraints, to describe a phenomena of the Hamiltonian localization
of classical mechanics. To achieve this result  it was necessary to
change the standard view on gauge fixing functions of the BRST-BFV
approach allowing them to depend on the variables of the auxiliary
sectors rather than  on the initial phase space variables. Then,
using the fact of the gauge independence of the partition function
constructed by the rules of the BRST-BFV method,  an application of
the unitary limit procedure allowed one to make the Lagrange
multipliers and the antighost variables dynamically - passive. As
the last step in our proof of the Hamiltonian localization,  it was
a special choice of the gauge fixing functions in the BRST-BFV
formalism.

Now, we would like to make a backward step as to explain our
understanding of what is the meaning of minimal Hamiltonian
delocalization. Let us modify slightly the unitary gauge
(\ref{eq.15}),
\beq
\label{eq.3.1}
 \chi^{A} = \Lambda^{A} - P_{A} \omega^{BA} + \frac{1}{2} g^{AB}
\Pi_{B}, 
\eeq
where $g^{AB}$ is an invertible matrix with the property of dual
antisymmetry,
\beq
\label{eq.3.2}
g^{AB} = - g^{BA} (-1)^{ ( \varepsilon_{A} + 1 ) ( \varepsilon_{B} + 1 )}. 
\eeq
Write down the gauge fixing term in (\ref{eq.11}),
with the gauge (\ref{eq.3.1}) substituted for (\ref{eq.15}),
\beq
\label{eq.3.3}
\Pi_{A} \chi^{A} = \Pi_{A} ( \Lambda^{A} - P_{B}\; \omega^{BA} ) +
 \frac{1}{2} \Pi_{A}\; g^{AB} \;\Pi_{B}.         
\eeq
With nonzero invertible $g^{AB}$, the Lagrange multipliers $\Pi_{A}$
enter quadratically (\ref{eq.3.3}), so that in the functional path integral,
in the unitary limit, the
$\Pi_{A}$ integration yields a Gaussian distribution of the unitary gauges
(\ref{eq.15}), instead of the delta - functional of (\ref{eq.15}).
That is actually what we mean when saying about the minimal Hamiltonian delocalization.

Finally, let us discuss another important aspect of the
construction. So far, we assumed
the original symplectic metric $\omega_{AB}$ as being constant. Now,
let us allow the metric be dependent of the original phase variables $Z^{A}$, although
satisfying the Jacobi relation,
\beq
\label{eq.3.4}
\omega^{AB}\; \overleftarrow{\pa_D}\; \omega^{DC} (-1)^{ \varepsilon_{A}
\varepsilon_{C} } + {\rm cycle}( A, B, C ) = 0.           
\eeq In order to have the original equations of motion be derivable
from the action of the form (\ref{eq.1}), we have to modify the
latter as follows (for details see \cite{BF})
\beq
\label{eq.3.5}
\frac{1}{2} \;\omega_{AB}\quad
\rightarrow \quad \bar{ \omega }_{AB} = ( N + 2 )^{-1}
\omega_{AB}, \quad N = Z^{A} \pa_{A} .     
\eeq
If one looks at the formula (\ref{eq.17}), it becomes clear
that the $\bar{C_{A}}$, $C^{B}$ integrations yield the correct
Jacobian only if the operator $\overleftarrow{\pa}_{B}$ applies both
to $H$ and to $\omega^{AC}$. On the other hand, our above arguments
are only capable to justify applying the operator
$\overleftarrow{\pa}_{B}$ to the $H$ only. To resolve that problem,
we should modify the gauge Fermion  $\Psi$  (\ref{eq.9}), (\ref{eq.15}) as
\beq
\label{eq.3.6}
\Psi \;  \rightarrow \; \Psi  +  \frac{ 1 }{ 2 } ( -1 )^{ \varepsilon_{D} }
\bar{ \mathcal{ P} }_{ D } \bar{ \mathcal{ P} }_{ A } \omega^{ AD }
\overleftarrow{\partial_B } \; C^{ B }.    
\eeq
Then the unitarizing Hamiltonian $\mathcal{ H }$ (\ref{eq.8}), (\ref{eq.11})
modifies as
\beq
\label{eq.3.7}
\mathcal{ H }\; \rightarrow \;\mathcal{ H }  +
\bar{ \mathcal{ P } }_{ A }\omega^{ AD } \overleftarrow{\partial_B }\;
\overrightarrow{\partial_D} H C^{ B } ( -1 )^{ \varepsilon_{ B }
\varepsilon_{ D } }.   
\eeq
which results exactly in the formula (\ref{eq.17}) where in the right-hand
side, in the integrand,  in the
second term, the left derivative $\overleftarrow{\partial_B }$
applies both to $H$ and to  $\omega^{ AC }$.
Thereby, the $\bar{ C }, C$  integrations yield exactly the Jacobian
of the argument of the delta function
yielded via the $P$ integration in the first term in the integrand.

\section{Lagrangian localization in field-antifield formalism}
As we have known that the whole field-antifield formalism is derived from
the Hamiltonian BRST- BFV quantization scheme  \cite{FVh1,BVh}, it appears quite natural to expect
that there should be possible to describe the localization phenomenon just
in the language of the field - antifield formalism, as well.
Let us begin with  determining the set of anticanonical
pairs of the field- antifield phase variables,  with the values assigned of the Grassmann
parity, $\varepsilon$,
and ghost  number, ${\rm gh}$.  For general field variables, $\Phi^{A}$, and their
antifields, $\Phi^*_{A}$, we have
\beq
\label{eq.4.1}
\varepsilon( \Phi^{A} )  =  \varepsilon_{A},  \quad   \varepsilon( \Phi^*_{A} ) =
\varepsilon_{A} + 1,\quad {\rm gh}( \Phi^{A} ) +  {\rm gh}( \Phi^*_{A} ) = - 1.                    
\eeq
For any functions $F, G$ on the antisymplectic phase space, the antibracket
is defined by
\beq
\label{eq.4.2}
( F, G ) = F ( \overleftarrow{\pa_A}\;\!\overrightarrow{\pa }^A_{*} -
\overleftarrow{\pa}^A_{*} \;\!\overrightarrow{\pa_A} ) G,      
\eeq
where $\pa_{A}$ and $\pa^A_{*}$ is the $\Phi^{A}$ and $\Phi^*_{A}$ - derivative,
respectively.

Now, we proceed with the following specific set of anticanonical pairs.

1. Original  phase variables:  $\phi^{i}, \phi^*_{i}$,
\beq
\label{eq.4.3}
\varepsilon( \phi^{i} ) = \varepsilon_{i}, \quad \varepsilon( \phi^*_{i} ) =
\varepsilon_{i} + 1,\quad
{\rm gh}( \phi^{i} ) = 0,   \quad    {\rm gh}( \phi^*_{i} ) = - 1;     
\eeq

2. Nakanishi - Lautrup  phase variables:  $B_{i},  B^{*i}$,
\beq
\label{eq.4.4}
\varepsilon( B_{i} ) = \varepsilon_{i} , \quad   \varepsilon( B^{*i} )
=\varepsilon_{i} + 1,\quad
{\rm gh}( B_{i} ) = 0, \quad  {\rm gh} ( B^{*i}) = - 1;     
\eeq

3. Ghost phase variables: $C^{i},  C^*_{i}$,
\beq
\label{eq.4.5}
\varepsilon( C^{i} ) = \varepsilon_{i} + 1,     \quad \varepsilon( C^*_{i} ) =
\varepsilon_{i},\quad
{\rm gh}( C^{i} ) = 1,  \quad   {\rm gh}( C^*_{i} ) = - 2;     
\eeq

4. Antighost phase variables:  $\bar{C}_{i},  \bar{C}^{*i}$,
\beq
\label{eq.4.6}
\varepsilon( \bar{C}_{i} ) = \varepsilon_{i} + 1,  \quad   \varepsilon(
\bar{C}^{*i} ) = \varepsilon_{i},\quad
{\rm gh}( \bar{C}_{i} ) = - 1,  \quad  {\rm gh}( \bar{C}^{*i}  ) = 0.    
\eeq

Now, let $\mathcal{S}( \phi )$ be an original action whose Hessian,
\beq
\label{eq.4.7}
H_{ik} = \overrightarrow{\pa_i} \;\!\mathcal{S} \;\!\overleftarrow{\pa_k},     
\eeq
is invertible at the extremals.  Here $\pa_i$ is the $\phi^i$ - derivative.
 Next, we define the following classical
master - action,
\beq
\label{eq.4.8}
S = \phi^*_{i} C^{i} + B_{i} \bar{C}^{*i}, \quad \varepsilon( S ) = 0, \quad  {\rm gh}( S ) =
0,  \quad   ( S, S ) = 0.  
\eeq
The gauge - fixing Fermion has the form,
\beq
\label{eq.4.9}
\Psi = \bar{C}_{i} \chi^{i}, \quad \varepsilon(\Psi) = 1, \quad {\rm gh}( \Psi ) = - 1,
\eeq
where just the gauge - fixing function is defined by
\beq
\label{eq.4.10}
\chi^{i} = g^{ik} ( \pa_{k} \mathcal{ S } + \frac{1}{2} \kappa B_{k} ),\quad
\varepsilon( \chi^{i} ) =  \varepsilon_{i}, \quad {\rm gh}( \chi^{i} ) = 0,
\eeq
$\kappa$ is the delocalization parameter. Metric $g^{ik}$ is invertible and
allowed to depend on $\phi^{i}$, it has the dual anti-symmetry property,
\beq
\label{eq.4.11}
g^{ik} = - g^{ki} (- 1)^{ ( \varepsilon_{i} + 1 ) ( \varepsilon_{k} + 1 ) }.
\eeq
The quantum partition function  reads
\beq
\label{eq.4.12}
\mathcal{Z } = \int D\phi DB DC D\bar{C} \exp\left\{ \frac{i}{\hbar } S_{\Psi }\right\},
\eeq where the gauge - fixed master - action is
\beq
\label{eq.4.13}
S_{ \Psi }  =  S|_{ \phi^{*} = \pa \Psi, \;\; \bar{C}^{*} = \chi  } =
\bar{C}_{i} g^{ik} ( \overrightarrow{\pa_k} \mathcal{ S } +
\frac{1}{2} \kappa B_{k} ) \overleftarrow{\pa_j} \;\! C^{j} + B_{i}
g^{ik} ( \overrightarrow{\pa_k} \mathcal{ S } + \frac{1}{2}
\kappa B_{k} ).
\eeq
As the metric $g^{ik}$ and delocalization parameter $\kappa$ do enter via the
gauge fixing (\ref{eq.4.10}) only, the partition function is independent of them, due to
the standard general arguments \cite{BVh,BV,BV1,KT,BF}.  Formally, for $\kappa = 0$,  the  $B^{i}$
- integration yields the delta - function of the extremal of the action $\mathcal{ S }$,
while the $\bar{C}_{i}, C^{k}$ - integrations yield the corresponding  Jacobian correct.

The solution chosen (\ref{eq.4.8}) for the classical master - action is a very
specific one. It is a homogeneous linear function of the antifields, so that it vanishes
at zero values of the ones. There is no term in (\ref{eq.4.8}) of zero order in the
antifelds. The latter circumstance makes it possible to choose the  unity matrix to stand
for the "gauge generator".

\section*{Acknowledgments}
\noindent
 I. A. Batalin would like  to thank Klaus Bering of Masaryk
University for interesting discussions. The work of I. A. Batalin is
supported in part by the RFBR grants 14-01-00489 and 14-02-01171.
 The work of P. M. Lavrov is supported by the Ministry of Education and Science of
Russian Federation, grant project 2014/387/122.

\end{document}